\documentclass{aa}
\usepackage{graphicx}
\usepackage{txfonts}
\begin{document}
   \title{Matched filters for source detection in the Poissonian noise regime}

   \author{I. M. Stewart
          \inst{1}
          }

   \offprints{I. M. Stewart}

   \institute{Jodrell Bank Observatory, University of Manchester,
              Macclesfield SK11 9DL, United Kingdom\\
              \email{ims@jb.man.ac.uk}
             }

   \date{Received January 0, 0000; accepted January 0, 0000}

   \abstract{
   A procedure is described for estimating an optimum kernel for the detection by convolution of signals among Poissonian noise. The technique is applied to the detection of x-ray point sources in XMM-Newton data, and is shown to yield an improvement in detection sensitivity of up to 60\% over the sliding-box method used in the creation of the 1XMM catalog.

   }

   \maketitle

%++++++++++++++++++++++++++++++++++++++++++++++++++++++++++++
\section{Introduction}

Over the past few years, CCD cameras on satellite observatories such as ROSAT, ASCA, Chandra and most recently XMM-Newton have generated high-resolution digital images of the x-ray sky which are characterized by relatively faint background (for example, background fluxes of less than 1 count per CCD pixel are often seen in typical-duration XMM-Newton exposures). The number of events per pixel follows a Poisson probability distribution which, at such low flux levels, deviates markedly from its Gaussian bright-end limit.

It seems likely that a significant part of the (non-instrumental) x-ray background is comprised of point sources too faint to be distinguished from one another by present techniques (Mushotzky et al \cite{mushotzky}, Hasinger et al \cite{hasinger}). The desire to characterise this population of sources is one reason for attempting to push the sensitivity of x-ray point source detection to the lowest limits allowed by the data.

Several source-detection procedures have been applied to x-ray images, eg: sliding-box (DePonte \& Primini \cite{pros}; Dobrzycki et al \cite{celldetect}; see also task documentation for the exsas task \emph{detect} and the SAS task \emph{eboxdetect}); wavelet (Damiani et al \cite{damiani}; Starck \& Pierre \cite{starck_and_pierre}; Pierre et al \cite{lss}; see also task documentation for the CIAO task \emph{wavdetect} and the SAS task \emph{ewavelet}); maximum-likelihood PSF fitting (Cruddace et al \cite{cruddace}; Boese and Doebereiner \cite{ml_fitting}); and Voronoi tessellation (Ebeling \& Wiedenmann \cite{voronoi}). There are also occasional references to use of a `matched filter' technique, but in the x-ray sphere at least this appears to consist just of convolution by the Point Spread Function or PSF (Vikhlinin et al \cite{vikhlinin}; Alexander et al \cite{chandra_matched}). A similar technique has also been applied to ASCA data (Ueda et al \cite{ueda}). As is shown in the present paper, the PSF approaches a true matched filter in the limit of white Gaussian noise but is sub-optimal at low levels of background.

A useful review of source-detection procedures as applied to x-ray images can be found in Valtchanov et al (\cite{valtchanov}).

Many of these techniques include a step in which the raw image is subjected to a convolution, often as a first step in preparing a detection-likelihood map. Even PSF fitting, in simplified form, can be shown to be equivalent to a convolution (Stetson \cite{stetson}). Voronoi tessellation seems to be the only technique which cannot readily be understood in this form.

Extraction of source positions and detection likelihoods from the convolved image is in general not simple, because the characteristic width of the PSF of the mirror system is usually larger than the CCD pixel size. In a (raw) image where the PSF is resolved, neighbouring pixels are not statistically independent - detection of some source flux in one pixel makes it more likely to detect it in neighbouring pixels. However, if the null hypothesis (ie, that there are no sources in the field) is assumed, detected counts are just due to background. But the background (by definition) must be slowly varying over the scale of the PSF, otherwise it would be impossible to separate it from the sources; and in the limit of smooth background the event counts in the raw image \emph{are} statistically independent.

On the other hand, as has already been pointed out, a significant fraction of the cosmological x-ray background (perhaps approaching 100\% at energies above 1 keV) actually consists of sources. This seems in direct contradiction to the null hypothesis. It is shown however in appendix \ref{confusion} that the bulk of these background sources must be very faint, and that for x-ray telescopes of presently achievable effective areas, one can model the net contribution of this source population by a relatively sparse population of brighter sources superimposed upon a smooth background. I conclude therefore that it is acceptable to test the null hypothesis, thus in effect to detect sources, on a pixel-by-pixel basis.

Some source-detection chains which assume the null hypothesis have adopted the following simplified scheme:

\begin{enumerate}
  \item Determination of the expectation value of background at all pixels;

  \item Convolution of the raw image;

  \item Calculation of a null-hypothesis likelihood map, which is just the likelihood, given the assumed background, of each value of the convolved image arising from it by chance;

  \item Location of sources by centroiding of troughs in the likelihood map; dealing with confused sources; parameter measurement, etc.
\end{enumerate}

The purpose of the present paper is to describe methods of calculating and applying a `matched filter', that is a convolver which is optimized for the detection, on a pixel-by-pixel basis, of sources of a given PSF against a given background. Only steps 2 and 3 of the above sequence are considered - ie, it is assumed that a good estimate of the background is available, and that source centroiding, confusion resolution etc in the likelihood map may be independently optimized.

The matched-filter detection scheme is compared with the sliding-box technique, specifically that used in the construction of the 1XMM catalog (Watson et al \cite{1xmm_paper}). The sliding box technique is arguably the simplest and most widely used of the detection techniques and thus provides a convenient baseline for comparison.

%++++++++++++++++++++++++++++++++++++++++++++++++++++++++++++
\section{Linear signal detection}
%oooooooooooooooooooooooooooooooooooooooooooooooooooooooooooo
\subsection{General} \label{sd_gen}

Suppose we have a parent function $C$ which is a function of some independent variables $\vec{x}$ (eg time, position, energy) and which comprises a background $B$ and signal $S$ such that

\begin{equation} \label{equ0}
  C(\vec{x}) = B(\vec{x}) + \alpha S(\vec{x}-\vec{x}_0)
\end{equation}
where
\begin{displaymath}
  \int_{-\infty}^{\infty} d\vec{x} \ S(\vec{x}) = 1,
\end{displaymath}

\noindent
$\alpha$ is the signal amplitude and $\vec{x}_0$ is a reference point which serves to locate the signal in $\vec{x}$ space. Let each experimental measurement of $C$ return a random variable $c$ which is distributed according to a probability distribution with an expectation value $\langle c \rangle = C$. A common problem in signal detection occurs when one one knows (or can estimate) $B(\vec{x})$ and $S(\vec{x})$ \emph{a priori} and one wishes to calculate the most likely values of $\alpha$ and $\vec{x}_0$ from a set of samples $c_i$ of $C$ at different values of $x$.\footnote{For simplicity I'll assume in this section that there is only one signal $S$ to be found, although, since convolution filtering is a linear process, in principle there is no difficulty in detecting many superposed signals provided they have sufficiently different values of $\vec{x}_0$.} In the case of point source detection, $\vec{x}$ is a two-coordinate vector which specifies position in the focal plane of a camera, $S$ is the point spread function (PSF) of the camera, and the samples $c_{i,j}$ are measurements of flux made on a pixel grid in the focal plane.

As outlined in the introduction, to assess whether there is any signal present in a given channel $i$ one calculates the probability of the parent background $B_i$ alone generating either the observed count $c_i$ or any value higher than this. This is called the probability of the null hypothesis ($P_\mathrm{null}$). A signal is judged to be `detected' in channel $i$ if the null probability of the measured $c_i$ falls below a previously selected cutoff value.

It is very often the case in practice that $S$ extends over several channels. In this case it is usually possible to improve the signal-to-noise ratio in at least one of the channels spanned by the signal, hence the detection sensitivity, by performing a weighted sum of the counts measured over several adjacent channels. In the XMM-Newton case which we are going to consider, $\vec{x}$ extends along spatial dimensions $x$ and $y$ and energy dimension $E$. The weighted sum is to be computed for each spatial pixel $(i,j)$, the eventual aim being to produce a map of the null-hypothesis likelihood at each $(i,j)$. For computational purposes the spatial sum is most conveniently expressed as a convolution; the actual expression employed to calculate the weighted sum for each pixel is therefore

\begin{equation} \label{true_weighted_sum}
  c^\prime_{i,j} = \sum_{p=-M}^{M} \sum_{q=-M}^{M} \sum_{k=1}^{N_\mathrm{bands}} w_{p,q,k} c_{i-p,j-q,k}.
\end{equation}

\noindent
For present purposes however it is unnecessary to retain such complication. As far as the statistical analysis goes, equation \ref{true_weighted_sum} at any given spatial pixel $(i,j)$ is simply a weighted sum of random variates:

\begin{equation} \label{weighted_sum}
  c^\prime = \sum_{i=1}^{N} w_i c_i.
\end{equation}

\noindent
It is also convenient for the present to assume that the signal is bracketted between $i=1$ and $i=N$, or, in other words, that the sampled parent function $C_i$ is given by

\begin{displaymath}
  C_i = B_i + \alpha S_i.
\end{displaymath}

\noindent
This is equivalent to assuming that a source is centred on pixel $(i,j)$ of equation \ref{true_weighted_sum}.

The standard deviation $\sigma^\prime$ of $c^\prime$ as given in equation \ref{weighted_sum} is estimated from the usual error propagation relations to be

\begin{equation} \label{std_dev}
  \sigma^{\prime 2} = \sum_{i=1}^N w^2_i \sigma^2_i
\end{equation}

\noindent
where $\sigma_i$ is the standard deviation of $c_i$.

Clearly, whatever weight scheme is adopted, the stronger a signal is (ie, the larger the signal amplitude $\alpha$), the higher the probability that it will be detected (the smaller the value of $P_\mathrm{null}$). Ideally we would like to be able to calculate some cutoff value of $\alpha$, above which the signal definitely would be detected, and below which it definitely would not. It is however not possible to do this, for the reason that the any non-trivial function $c^\prime$ of the detected counts must itself be a random variable. A given value of $\alpha$ will not always give rise to the same value of $c^\prime$, and any value of $\alpha$ will produce any chosen value of $c^\prime$ within a sufficiently large ensemble. To get around the problem and so to permit the comparison of different choices of weights it is convenient to define a `counts amplitude' $\beta$ such that

\begin{equation} \label{beta_def}
  \beta  = \frac{c^\prime - B^\prime}{S^\prime}
\end{equation}

\noindent
where the significance of the primes here is that, for any function $f$,

\begin{displaymath}
  f^\prime = \sum_{i=1}^N w_i f_i.
\end{displaymath}

\noindent
It is not hard to show that $\langle \beta \rangle  = \alpha$. 

Since any given null-hypothesis probability is associated with a definite value of the weighted sum of counts $c^\prime$, it is also therefore unambiguously associated with a definite value of $\beta$. This allows us to define the detection sensitivity $\beta_\mathrm{det}$ of any convolution as that value of $\beta$ which is associated with the chosen detection-cutoff value of $P_\mathrm{null}$.

Clearly it is also the case that, for a given $B_i$ and $S_i$, there will be a set of weights (not necessarily unique) which gives maximum sensitivity, ie the smallest possible value of $\beta_\mathrm{det}$.

%oooooooooooooooooooooooooooooooooooooooooooooooooooooooooooo
\subsection{The Gaussian case}

For purposes of comparison I'll briefly reprise the well-studied case where the probability distribution of $c_i$ for each $i$ is Gaussian, with constant $\sigma_i = \sigma$. In this case any weighted sum returns $c^\prime$ values which also have a Gaussian distribution, with

\begin{displaymath}
  \sigma^{\prime 2} = \sigma^2 \sum_{i=1}^N w^2_i.
\end{displaymath}

\noindent
One can calculate the signal-to-noise ratio $snr$ as follows:

\begin{displaymath}
  snr = \frac{c^\prime - B^\prime}{\sigma^\prime}.
\end{displaymath}

\noindent
The null-hypothesis probability is then given by

\begin{displaymath}
  P_\mathrm{null}(snr) = 0.5 \left( 1 - \mathrm{erf} \left[ snr/\sqrt{2} \right] \right).
\end{displaymath}

\noindent
By differentiating $snr$ with respect to the weights $w_i$ and equating each of the $N$ resulting derivatives to zero one arrives at the well-known result that the optimum set of weights is proportional to the signal itself: ie

\begin{displaymath}
  w_i = k S_i \ \forall \ i,
\end{displaymath}

\noindent
where $k$ is some non-zero constant.

%oooooooooooooooooooooooooooooooooooooooooooooooooooooooooooo
\subsection{Weighted-Poisson case with a single weight}

In the case where the probability distribution of the observed values $c_i$ is Poissonian, the optimum weights are not so easy to come by, and in general will depend in a non-trivial fashion on the level of background $B_i$.

Consider first the simplest case, in which there is only one weight ($N=1$ in equation \ref{weighted_sum}). The solution in this case is trivial, but provides a useful mathematical template for the more difficult case in which $N > 1$. Let $c$ be a random (necessarily integer) variable which has a Poisson probability distribution. A second variable $c^\prime$ which is just $c$ multiplied by a single weight $w$ retains the Poisson probability distribution

\begin{equation} \label{pnull_basic}
  p(c^\prime) = \frac{\nu^c e^{-\nu}}{c!},
\end{equation}

\noindent
but with $\nu$ now given by

\begin{displaymath}
  \nu = \langle c \rangle = \frac{\langle c^\prime \rangle }{w}.
\end{displaymath}

In the null hypothesis, $\langle c \rangle = B$ and thus $\langle c^\prime \rangle  = wB = B^\prime$. If we extrapolate from the unscaled Poissonian case it is also clear that the null-hypothesis probability $P_\mathrm{null}$ is given by

\begin{equation} \label{equ1}
  P_\mathrm{null}(c^\prime) = 1-Q \left(\frac{c^\prime}{w}, \frac{\langle c^\prime \rangle }{w} \right) = 1-Q \left(\frac{c^\prime}{w}, \frac{B^\prime}{w} \right)
\end{equation}

\noindent
where $Q$ is the (complementary) incomplete gamma function, defined by

\begin{displaymath}
  Q(a,x) = \frac{1}{\Gamma(a)} \int_x^{\infty} dt \, e^{-t} t^{a-1}.
\end{displaymath}
$\Gamma$ here represents the gamma function.

%oooooooooooooooooooooooooooooooooooooooooooooooooooooooooooo
\subsection{Weighted-Poisson case with $N>1$ weights} \label{many_n_theory}

Formally speaking, we are now no longer in the Poissonian regime, since a weighted sum of two or more Poissonian variates does not itself in general have a Poissonian probability distribution. I have not been able to find a closed-form expression for the probability density function in this case. However, two empirical approximations are presented in the present subsection.

%............................................................
\subsubsection{The Fay and Feuer approximation} \label{ff_approx}
Fay and Feuer (\cite{fay_and_feuer}) suggested on heuristic grounds that the null-hypothesis probability in this general case might be given to a good approximation by equation \ref{equ1} with $w$ replaced by an appropriate equivalent weight $w_\mathrm{equiv}$. Their prescription for $w_\mathrm{equiv}$,

\begin{displaymath}
  w_\mathrm{equiv} = \frac{\sigma^{\prime 2}}{\langle c^\prime \rangle}.
\end{displaymath}

\noindent
where

\begin{displaymath}
  \langle c^\prime \rangle  = \sum_{i=1}^N w_i \langle c_i \rangle 
\end{displaymath}

\noindent
and, from equation \ref{std_dev}

\begin{displaymath}
  \sigma^{\prime 2} = \sum_{i=1}^N w_i^2 \langle c_i \rangle ,
\end{displaymath}

\noindent
is obtained essentially by equating respectively the first and second moments of the single-weight and many-weight probability functions.

\begin{figure}
   \centering
      \resizebox{\hsize}{!}{\includegraphics[angle=-90]{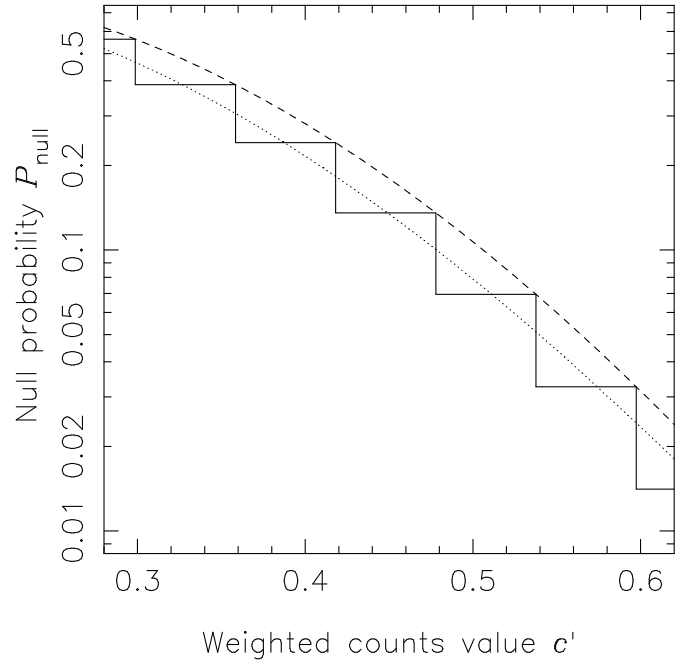}}
      \caption{A comparison between various weighted-Poisson integrated probability functions. The dotted line shows the results of a Monte Carlo experiment in which $c^\prime$ was formed as a weighted sum of 25 independant Poisson variates with an expectation value $B$ of 0.3. The value of the equivalent single weight $w_\mathrm{equiv} = \sigma^{\prime 2} / B^\prime$ was $5.97 \times 10^{-2}$. The remaining two curves refer to a single Poisson variable $c^\prime$ of the same expectation value 0.3, weighted by $w_\mathrm{equiv}$. The solid stepped line represents the true probability $P_\mathrm{null}(c^\prime)$ of obtaining a measured value equal to or greater than $c^\prime$ in the single-weight case. The dashed line is the envelope function to this $P_\mathrm{null}(c^\prime)$, as given by equation \ref{equ1}.
              }
         \label{nullHypProbs}
   \end{figure}

In fact when one compares the integrated $P_\mathrm{null}(c^\prime)$ function derived from the Fay and Feuer approximation against Monte Carlo data (see figure \ref{nullHypProbs}), one sees that the Fay and Feuer curve seems to be displaced too far to high $c^\prime$. Essentially this is because equation \ref{equ1} defines a continuous \emph{envelope} function to the actual discontinuous, stepped $P_\mathrm{null}$ of a single Poisson variate. Consider now a weighted sum of $N$ Poisson variates, but with all the weights having the single value $w$: here the sum itself remains a Poisson variate, in which case the true integrated probability distribution remains coarsely stepped as in figure \ref{nullHypProbs}, the envelope to this being exactly given by equation \ref{equ1}, with $w_\mathrm{equiv} = w$. The coarseness is because there are no combinations of $c_i$ which can produce any value of $c^\prime$ other than $c^\prime = jw$ for integer $j$; the steps in the integrated probability distribution $P_\mathrm{null}$ are thus of width $w$. If we now allow the weights to be randomly perturbed by small amounts, the effect is to smear out the steps: ie, a range of values of $c^\prime$ close to $iw$ now become possible. Where the weights $w_i$ are allowed to become entirely random and independent (and are sufficently numerous), the coarse steps disappear entirely\footnote{Note that the true probability curve in the case that $N > 1$ still falls in stepwise fashion, since the possible values of $c^\prime$ in this case are still discrete; it is just that the steps are much more finely spaced, since the spacing between possible $c^\prime$ values (and thus also between steps) decreases on average at a rate proportional to $(c^\prime)^{1-N}$.}, and the probability curve appears to steer a middle course (eg the dotted line in figure \ref{nullHypProbs}) through the coarse steps of the single-weighted Poisson distribution of the same equivalent weight. It seems clear then that the approximation formula suggested by Fay and Feuer could be made more closely applicable to the general case by shifting it towards lower $c^\prime$ values by an amount $0.5 w_\mathrm{equiv}$. The resulting formula is

\begin{equation} \label{equ3}
  P_\mathrm{null}(c^\prime) = 1-Q \left( \frac{c^\prime}{w_\mathrm{equiv}} + \frac{1}{2}, \frac{B^\prime}{w_\mathrm{equiv}} \right).
\end{equation}

Figure \ref{nullHypProbs} shows an example of a $P_\mathrm{null}$ distribution derived from a Monte Carlo exercise (dotted line). The Monte Carlo ensemble consisted of $10^6$ values of the weighted sum $c^\prime$. 25 random weights $w_i$ were chosen before the start of the exercise: these initially had a uniform probability distribution between 0 and 1 but were then normalized so that they summed to 1. (The only effect of this normalization is to change the x-axis scale.) For each member of the ensemble, a Poisson-random integer $c_i$ was generated for each of the 25 bins, using a constant background value $B = 0.3$ counts per bin as the Poisson expectation value. The weighted sum $c^\prime$ of these 25 random values was made and added to the ensemble. The cumulative probability was then formed by summing, from high towards low values of $c^\prime$, a 100-bin, normalized histogram of the ensemble values. The modified Fay and Feuer curve (equation \ref{equ3}) has not been plotted, but its path can be easily visualised by mentally shifting the dashed curve to the left by half the width of the coarse steps.

Empirical tests with randomly-chosen weights and background suggest that equation \ref{equ3} is a reasonable fit to actual distributions of $c^\prime$ for values of null probability greater than about $10^{-2}$. Better fits were observed when the distribution of both weights and background was even, and $B^\prime$ was greater than about 0.1. At values of $P_\mathrm{null}$ smaller than about $10^{-2}$, test data appear to diverge from equation \ref{equ3}, such that the actual null probability at a given value of $c^\prime$ is larger than the predicted value. The divergence appears to worsen at low values of background, or if the weights are not very homogeneously distributed. Comparison of the respective expressions for the third moments $\mu_3 = \langle (c^\prime)^3 \rangle$ of the single-weight and multiple-weight distributions shows that $\mu_{3,\mathrm{multi}}$ is always greater than $\mu_{3,\mathrm{single}}$ for positive, nonequal $w_i$. It is therefore almost certain that all the higher moments differ as well. If this is the case a divergence at high $c^\prime$ between the many- and single-weight distributions is to be expected.

Attempts to fit a function of the form of equation \ref{equ3} to test data were not satisfactory - that is, the best fit was still clearly not a good approximation to the parent distribution at low $P_\mathrm{null}$. Since source detection is necessarily concerned with low values of the probability of the null hypothesis, it is desirable to find a better approximation in this region.

%............................................................
\subsubsection{The $\chi^2$ approximation} \label{chi2_approx}

\begin{figure}
   \centering
      \resizebox{\hsize}{!}{\includegraphics[angle=-90]{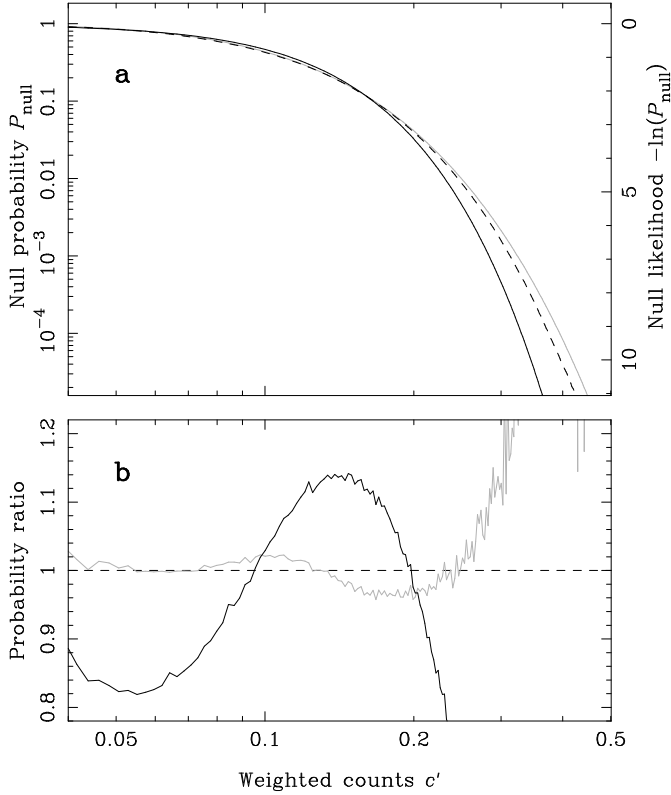}}
      \caption{An example plot comparing approximate expressions for the null probability $P_\mathrm{null}$ for a weighted sum $c^\prime$ of Poisson variates at lower values of $P_\mathrm{null}$ than were explored in figure \ref{nullHypProbs}. \textbf{a)} The dashed line plots the result of a Monte Carlo experiment in which the weights were optimized for detecting sources in XMM-Newton EPIC images, as described in section \ref{matched_1_band}. The solid black line gives the modified Fay and Feuer approximation of equation \ref{equ3}; the solid grey line, the $\chi^2$ approximation of equation \ref{chi_for_many_W}. \textbf{b)} This figure, with the same horizontal scale as \textbf{a}, shows ratios between probability densities $p$. The solid black and grey lines show respectively the Fay and Feuer and $\chi^2$ $p$ divided by the Monte Carlo $p$.
              }
         \label{low_P_null}
   \end{figure}

The $\chi^2$-like integrated probability distribution

\begin{equation} \label{chi_for_many_W}
  P_\mathrm{null}(c^\prime) = Q \left(\frac{B^\prime}{w_\mathrm{equiv}}, \frac{c^\prime}{w_\mathrm{equiv}} \right)
\end{equation}
proved to be an acceptable approximation down to at least $P_\mathrm{null} = 10^{-5}$ over a wide range of background values. An example is shown in figure \ref{low_P_null}.

The Monte Carlo simulation used to generate the results of figure \ref{low_P_null} was similar to that of figure \ref{nullHypProbs}. It consisted of an ensemble of $10^7$ weighted sums. The weights in this case constituted a matched filter for detection of sources in images with an average background of 0.1 counts/pixel, the source shape being the on-axis point spread function at 1.25 keV of the XMM-Newton EPIC PN camera.

It might be possible to improve the fit of equation \ref{chi_for_many_W} by choosing a different value of $w_\mathrm{equiv}$, but this possibility has not been explored.

It is not at present known why the $\chi^2$ cumulative distribution given in equation \ref{chi_for_many_W} seems to provide such a good model of the weighted-sum $P_\mathrm{null}$.

%............................................................
\subsubsection{Consequences of divergence of an approximation} \label{consequences}

Suppose one inverts a formula such as equation \ref{equ3} in an attempt to deduce the detection sensitivity at a given value of $P_\mathrm{null}$. What in particular goes wrong if the formula is not a good approximation to the true probability distribution? The answer is that the returned sensitivity value is correct, but the assumed $P_\mathrm{null}$, which controls the number of false positives expected, will be incorrect. In the example of figure \ref{low_P_null}\textbf{a}, cutting the source list at an apparent null likelihood of 8 will yield a sensitivity of about 0.31 weighted counts if the Fay and Feuer expression is used. The true null likelihood at that value of $c^\prime$ is that of the Monte Carlo, ie about 6.7. This means that about 3.7 ($=e^{8-6.7}$) times more false positives will be encountered than expected. The $\chi^2$ formula on the other hand diverges from the Monte Carlo data in the other direction. Use of this formula gives a conservative result, being apparently slightly less sensitive at $c^\prime=0.36$, with the true null likelihood cutoff of about 8.6 yielding approximately 0.55 fewer false positives than expected. In order to get something like the desired rate of false detections, one must skew the apparent null likelihood cutoff, to about 10 in the Fay and Feuer case and 7.5 in the $\chi^2$ case. The sensitivity in both cases will then be `correct' at about 0.34 weighted counts.

%oooooooooooooooooooooooooooooooooooooooooooooooooooooooooooo
\subsection{Optimization of the weights} \label{optimization}

As described in section \ref{sd_gen}, for a given background $B$ and signal template $S$, any given value of the weighted sum of counts $c^\prime$ is associated with a unique value of the null-hypothesis probability $P_\mathrm{null}$. In the preceding subsection two equations (\ref{equ3} and \ref{chi_for_many_W}) were given which approximate this relation for the situation in which the observed data are random Poisson variates. In order to calculate optimum weights in this case we must first invert an equation of this form to obtain $c^\prime$, then invert equation \ref{beta_def} to obtain the counts amplitude $\beta$.

Inversion of either equation \ref{equ3} or \ref{chi_for_many_W} to obtain $c^\prime$ involves an inversion of the incomplete gamma function $Q$. For the sake of practicality let us define the two inverse functions $Q_1^{-1}$ and $Q_2^{-1}$ as follows: if
\begin{displaymath}
  P = 1-Q(a,x),
\end{displaymath}
then let
\begin{displaymath}
  a = Q_1^{-1}(1-P,x).
\end{displaymath}
and
\begin{displaymath}
  x = Q_2^{-1}(1-P,a).
\end{displaymath}

\noindent
No closed-form expressions for either $Q_1^{-1}$ or $Q_2^{-1}$ seem to be known; for present purposes I have performed the inversions numerically by means of a Ridders-method routine (Ridders \cite{ridders}) as given in Press et al (\cite{numerical_recipes}).

After insertion of the appropriate inverse gamma function, the counts amplitude $\beta$ is thus given for the modified Fay and Feuer approximation by

\begin{equation} \label{beta_def_ff}
  \beta = \frac{w_\mathrm{equiv} \left[ Q_1^{-1} \left( 1-P_\mathrm{null}, \frac{B^\prime}{w_\mathrm{equiv}} \right) - \frac{1}{2} \right] - B^\prime}{S^\prime}
\end{equation}

\noindent
and for the $\chi^2$ approximation by

\begin{equation} \label{beta_def_chi2}
  \beta = \frac{w_\mathrm{equiv} \left[ Q_2^{-1} \left( P_\mathrm{null}, \frac{B^\prime}{w_\mathrm{equiv}} \right) \right] - B^\prime}{S^\prime}.
\end{equation}

The inputs to these formulae are (i) the set of weights, (ii) the value of $P_\mathrm{null}$, and (iii) the background and signal-shape information. Only the first two of these are under our control. Before we may begin to seek for the optimal set of weights, we must choose a value of $P_\mathrm{null}$. The value we choose should be governed by the maximum fraction of false detections we are prepared to tolerate. The desired cutoff value $P_\mathrm{det}$ of null probability can be obtained from the maximum acceptable number $n_\mathrm{false}$ of false detections by dividing by the number of `beams' in a typical image, which is just the image solid angle divided by the `beam' solid angle. In the present case, in which images are convolved before the null hypothesis is tested, the beam solid angle must be some kind of equivalent solid angle of the appropriately convolved PSF. For XMM-Newton at least, where the shape of the PSF varies across the field of view, its average value is probably best estimated via Monte Carlo trials. If however, for the sake of obtaining at least a rough approximation to the ratio between $P_\mathrm{det}$ and $n_\mathrm{false}$, we use the value $2.34 \times 10^{-5}$ deg$^2$ calculated in appendix \ref{confusion} for the equivalent solid angle of the on-axis EPIC PN PSF as a lower limit to the beam $\Omega$, we find that a null-probability cutoff of $\exp(-8.0)$, the value used in the making of the 1XMM catalog, corresponds to at most 2 expected false detections per image.

Let us define $\beta_\mathrm{det}$ as the counts amplitude $\beta$ which, for a given set of weights, corresponds to the chosen value of $P_\mathrm{det}$. $\beta_\mathrm{det}$ can be viewed as the amplitude of a source which is just detectable under these conditions. The optimum weights are then clearly those which yield the smallest value of $\beta_\mathrm{det}$.

In the present study, Powell's direction-set method as modified by Press et al (Press et al \cite{numerical_recipes}, chapter 10.5) was used to optimize sets of weights by minimizing $\beta_\mathrm{det}$, as defined by either equation \ref{beta_def_ff} or \ref{beta_def_chi2}, as a function of the weights.

%++++++++++++++++++++++++++++++++++++++++++++++++++++++++++++
\section{Detection of x-ray point sources in XMM-Newton data}

%oooooooooooooooooooooooooooooooooooooooooooooooooooooooooooo
%\subsection{The x-ray background} \label{confusion}

%oooooooooooooooooooooooooooooooooooooooooooooooooooooooooooo
%\subsection{The XMM-Newton x-ray cameras} \label{cameras}
The EPIC x-ray cameras of XMM-Newton are described in Str\"uder et al (\cite{strueder}) and Turner et al (\cite{turner}). There are three cameras: the telescope in each case is similar but two of the CCD detectors comprise 7 chips of MOS type, whereas the third has 12 chips of pn composition. The `good' area of each of the three cameras occupies about 94\% (MOS) and 82\% (pn) of a 30$\arcmin$ diameter field of view. CCD pixel dimensions are 1.1$\arcsec$ square (MOS) and 4.1$\arcsec$ square (pn).

%oooooooooooooooooooooooooooooooooooooooooooooooooooooooooooo
\subsection{The source-detection strategy used for the 1XMM catalog} \label{1xmm_strategy}
For each exposure, images in sky coordinates were made in five separate energy bands. The images had square pixels of $4 \times 4$ arcsec dimension. Images were made by transforming the position on the detector of each selected x-ray event into sky coordinates, then binning up the events into the image pixels. The position of each event was dithered within the boundaries of the CCD pixel in which it was detected. Variations over time of the spacecraft attitude were also taken into account.

Source detection was performed on the five images in parallel. The source detection comprised a convolution and detection stage (steps 1 to 3 of the sequence described in the introduction), followed by a source-parameterisation stage (step 4 in the sequence). The detection etc stage was performed by the XMM-Newton SAS task \emph{eboxdetect}, the parameterisation by \emph{emldetect}. Both stages involve the calculation of a detection likelihood; since the value calculated by \emph{emldetect} is arguably more sensitive than that of \emph{eboxdetect}, the ideal procedure is to run \emph{eboxdetect} with a deliberately low detection threshold, submit the resulting long list of source candidates to \emph{emldetect} and to then accept as genuine sources only those for which the \emph{emldetect} detection likelihood exceeded a second, more reasonable threshold. The \emph{eboxdetect} threshhold should not be so close to the \emph{emldetect} threshhold that the two selections interfere. In 1XMM practice there is some doubt as to whether the two detection threshholds were sufficiently far apart. For this reason, and because I don't understand enough of the \emph{emldetect} likelihood calculation to be able to replicate it, I have in the present paper only considered the sliding-box stage of the 1XMM detection procedure.

The first step of this procedure was to make maps, 1 per energy band, of the estimated background in each pixel. The five images and five background maps were then each convolved with a square, $5 \times 5$ array of unit values.\footnote{Two further rebinning and convolution steps, approximately equivalent to convolution respectively by $10 \times 10$ and $20 \times 20$ arrays, were also performed. However, the principal purpose of these extra steps was to detect extended sources: therefore they can safely be neglected for purposes of the present discussion.} In mathematical form this processing can be represented as follows:

\begin{displaymath}
  c^\prime_{i,j,k} = \sum_{p=-2}^2 \sum_{q=-2}^2 c_{i-p,j-q,k}
\end{displaymath}

\noindent
and

\begin{displaymath}
  B^\prime_{i,j,k} = \sum_{p=-2}^2 \sum_{q=-2}^2 B_{i-p,j-q,k}
\end{displaymath}

\noindent
where $i$ and $j$ indicate the position on the image pixel grid and $k$ refers to the energy band.

The overall null-hypothesis probability was calculated as follows. Firstly, because all the weights are equal to 1, $c^\prime$ remains a Poissonian variate; equation \ref{equ1} is therefore exact at permitted values of $c^\prime$, with

\begin{displaymath}
  w = w_\mathrm{equiv} = 1.
\end{displaymath}

\noindent
The probabilities $P_{i,j,k}$ returned by equation \ref{equ1} were converted to likelihoods (ie, negative logs of the probabilities). For each image pixel, a summed likelihood $L_{i,j}$ was then generated:

\begin{equation} \label{summed_like}
  L_{i,j} = -\sum_{k=1}^M \ln(P_{i,j,k}),
\end{equation}

\noindent
where $M$ is the number of energy bands (here 5). The probability distribution of $L$ is however open to some question. Bevington and Robinson (\cite{bevington}) show that a sum of the same form as equation \ref{summed_like} is distributed approximately as $\chi^2/2$ (up to an additive constant) with $\nu = M$ degrees of freedom. Wilks (\cite{wilks}) as cited in Cash (\cite{cash}) comes to similar conclusions. Perhaps for this reason the authors of the XMM-Newton SAS task \emph{eboxdetect}, which performed the calculation of detection likelihoods for the 1XMM catalog, used the following approximate formula for the integrated probability of the null hypothesis across 5 bands:

\begin{equation} \label{chi_null_prob}
  P_{\mathrm{null},i,j} = Q \left(5, L_{i,j} \right).
\end{equation}

Note however that the first term represents 1/2 the degrees of freedom, hence should be 2.5 not 5; also, the $P$ in equation \ref{summed_like} are integrated probabilities, not probability densities.

   \begin{figure}
   \centering
      \resizebox{\hsize}{!}{\includegraphics[angle=-90]{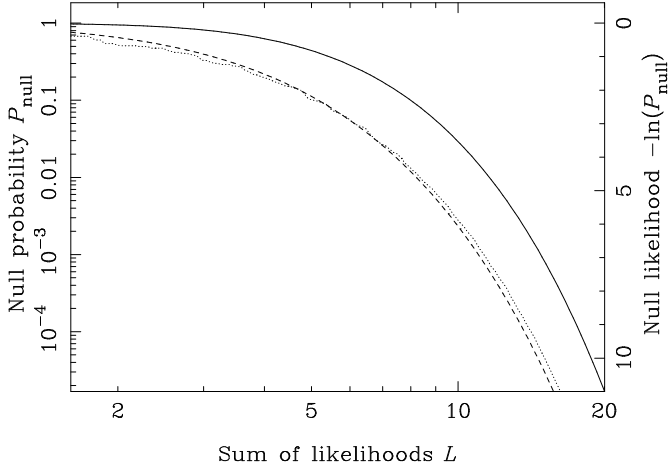}}
      \caption{Example probability curves relating to the 1XMM source-detection technique. The dotted line is the result of a Monte Carlo experiment (described in the text) which generated an ensemble of values of the summed likelihood $L$ described by equation \ref{summed_like}. The solid line gives the null-hypothesis probability predicted by equation \ref{chi_null_prob}. The dash-dot line is the function $Q(5.81/2, L)$, which was fitted to the Monte Carlo data by the procedure described in the text. Note that, for easier comparison, the vertical scale and the dynamic range of the horizontal scale are the same as those in figure \ref{low_P_null}.
              }
         \label{Chi2MonteCarlo}
   \end{figure}

Although a full analysis of the 1XMM source-detection technique is beyond the scope of the present paper, a Monte Carlo simulation of the statistical fluctuation of background was performed in order to check the accuracy of equation \ref{chi_null_prob}. An ensemble of $5\times 10^7$ values of $L$ was accumulated. The procedure for each member of the ensemble was as follows. For each of the 5 bands, a $5\times 5$ array of Poisson-random integers was generated. To calculate the expectation value $B_{i,j,k}$ for pixel $(i,j)$ in the $k$th band, the $k$th normalized background weight for the XMM PN camera, as listed in table \ref{table:1}, was multiplied by 0.1 counts/pixel. $c^\prime_k$ was calculated by summing the 25 counts values of the $k$th band, $B^\prime_k$ being of course simply equal to $25 \times B_{i,j,k}$. Likelihoods were then generated for each band by use of equation \ref{equ1}, and summed to give $L$.

The accumulated histogram of the resulting data, shown in figure \ref{Chi2MonteCarlo}, shows that there is a significant difference between the Monte Carlo results and the prediction of equation \ref{chi_null_prob}. Attempts were made to fit a variety of functions to the Monte Carlo data. One must of course fit to the raw histogram data, the adjacent channels of which are statistically independent, rather than to the integrated curve displayed in figure \ref{Chi2MonteCarlo}. It is not easy to fit a smooth function to the Monte Carlo data. Partly this is because the raw data are rather `noisy', particularly at low values of $L$, a phenomenon which arises because in this range the input counts values to each of the five channels are usually small integers which, when combined, give rise to an sparse distribution in the allowed resulting $L$ values. The effect of this can be seen in the jumpy nature of the accumulated Monte Carlo data shown in figure \ref{Chi2MonteCarlo}. It is also not straighforward to define a statistic to be minimized in order to generate the fit. A straightfoward sum of squared residuals, or even a chi squared sum (ie, sum of squared residuals, each divided by the variance in that channel), tends to result in the low-$L$ part of the data dominating the fit. This is undesirable if one is interested in minimizing false detections due to statistical fluctuations in background, in which case intermediate values of $L$ are more important. In the end I chose arbitrarily to minimize a sum of terms

\begin{displaymath}
  Z = \sum_i \frac{(f_i - y_i)^2}{\sigma_i^4}
\end{displaymath}

\noindent
where $f$ is the function to be fitted, $y$ represents the Monte Carlo data and $\sigma^4$ is the square of the variance. The fit was however restricted to values of $L$ less than 15, to avoid statistical noise in the Monte Carlo values at high $L$.

There is an infinity of functions one could choose to fit to the data, but in fact a good fit as shown was obtained simply by allowing the first term $\nu /2$ in the $Q$ function in equation \ref{chi_null_prob} to vary. The best fit, shown in figure \ref{Chi2MonteCarlo}, occurred at $\nu = 5.81$. This however suggests no obvious systematic correction to equation \ref{chi_null_prob} and, since I have at this time no better analysis of the problem to put forward, and since the unmodified equation \ref{chi_null_prob} was in fact used in the 1XMM source detection, I have retained it for comparison with the matched-filter method. It seems clear however that the rate of statistical false detections in 1XMM was probably lower than originally estimated, and thus that a lower detection cutoff could have been used in that survey without penalty. There appears to be no better way to correct estimates of the sensitivity of this method at the moment than by calibrating it via several Monte Carlos performed at different values of background (see eg section \ref{matched_at_5}).

A final point to note about the 1XMM detection procedure: it did not make use of the fact that, for any given exposure, images from more than one of the XMM-Newton x-ray cameras were usually available. All source detection was performed instead on a camera-by-camera basis. Clearly one would expect a technique which made use of the parallel information to yield an improvement in detection sensitivity.

%oooooooooooooooooooooooooooooooooooooooooooooooooooooooooooo
\subsection{Matched filters for XMM-Newton source detection}

There are three obvious ways to improve on the 1XMM source-detection procedure by use of matched filters (weighted sums). Firstly, taking one energy band at a time, since we know to a reasonable approximation the point spread function (PSF) of the XMM-Newton telescopes, we could choose in each energy band a set of weights optimized for detecting that shape of signal; secondly, we might do the same thing across the energy bands by the expedient of assuming a common spectrum for the sources; thirdly, we might in similar fashion add together images from the three x-ray cameras of XMM-Newton. Only the first two alternatives are explored in the present study.

%............................................................
\subsubsection{Examples of matched filters for 1 energy band} \label{matched_1_band}

We take a square array of dimension $2N+1$. Let us assume a parent function $C_{i,j}$ as follows:

\begin{displaymath}
  C_{i,j} = B_{i,j} + \alpha S_{i,j}, \ \forall \ i,j \ \mathrm{in} \ [-N,N]
\end{displaymath}
with the normalization condition
\begin{displaymath}
  \sum_{i=-N}^N \sum_{j=-N}^N S_{i,j} = 1.
\end{displaymath}

\noindent
For simplicity, the background $B$ is assumed to be constant across the array. Let $S_{i,j}$ be the PSF on the optic axis of the PN telescope of XMM-Newton, at an energy of 1.25 keV (the mean energy of band 2 as defined in the 1XMM catalog (Watson et al \cite{1xmm_paper})). The PSF model used in the present exercise was originally calculated via a ray-tracing approach (Gondoin et al \cite{gondoin}), and is the same that was used to determine positions and fluxes of the sources in 1XMM. In the present exercise the PSF was centred on the middle of the $(i,j)=(0,0)$ pixel.\footnote{In practice, the centre of the PSF for any real source can of course be located anywhere within the `central' pixel. A more rigorous procedure would take this into account. However, because it is not clear how best to do this, the simpler assumption has been made for the present study.}

We assume that we have an array of measured count values $c_{i,j}$ across the array, each of which is a random, Poisson variable, with $\langle c_{i,j} \rangle  = C_{i,j}$. We make a weighted sum of the $c_{i,j}$ as follows:

\begin{displaymath}
  c^\prime = \sum_{i=-N}^N \sum_{j=-N}^N w_{i,j} c_{-i,-j},
\end{displaymath}

\noindent
and likewise for $B^\prime$.

The two cases we want to compare are, firstly, the 1XMM-like case in which the $w_{i,j} = 1$ and $N = 2$; secondly the case in which the $w_{i,j}$ are optimized according to the procedure described in section \ref{optimization}. But before the second procedure can be used, a value of $N$ must be chosen. It is not hard to see that, in principle, the larger the $N$ the better. To see this, suppose we compare two convolvers: one optimised on a ($2N+1$)-square array, the other on a ($2(N-1)+1$)-square array. The optimized values of the small convolver can be thought of as a subset of the possible values of the large convolver; one just sets the extra ring of pixels to zero. So, if the smaller convolver were a better source-finder than the large, the optimisation routine would have set the outer pixel values to zero automatically, giving rise to the same sensitivity of detection with the large as with the small. Thus a small optimized convolver can never be better than a large one; and the only limiting factor to $N$ becomes computational practicality. A value of $N = 4$ for the optimized convolvers has been chosen as the largest value which can be processed in the 5-band case (see section \ref{matched_at_5}) in a reasonable time.

\begin{figure}
   \centering
      \resizebox{\hsize}{!}{\includegraphics[angle=-90]{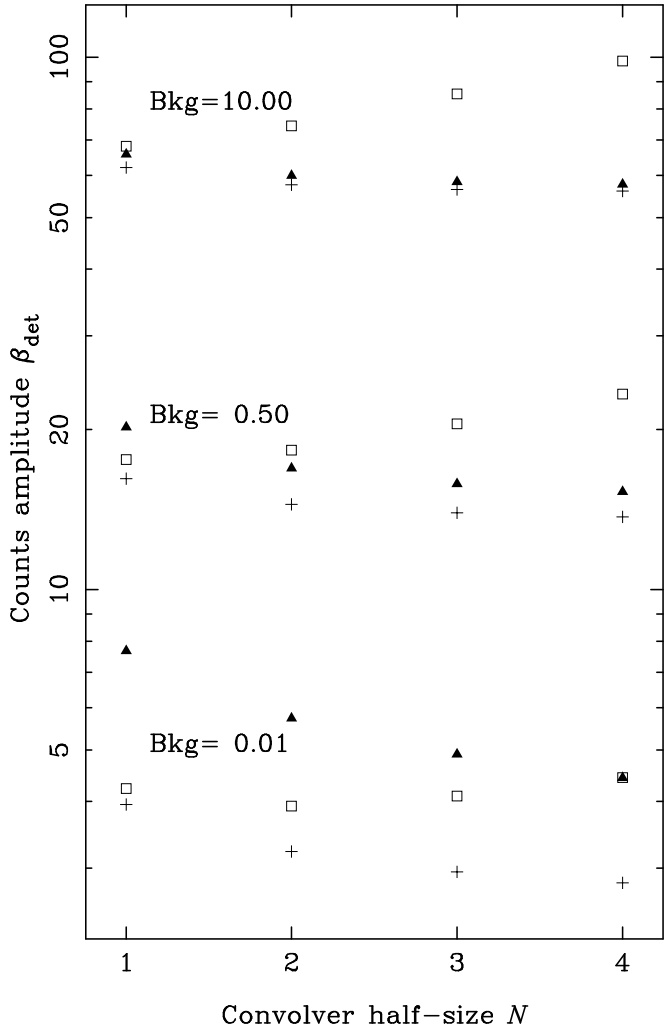}}
      \caption{Sensitivity of different convolvers as the array size is varied. The array side length is $2N+1$. The squares represent the 1XMM `box' convolver. The optimized convolver is represented by both crosses and filled triangles: the crosses show the estimates obtained by using the modified Fay and Feuer approximation (equation \ref{beta_def_ff}), whereas the triangles show the estimates from the chi-squared formula (equation \ref{beta_def_chi2}). Those cases in which the sensitivity of the optimized convolver appears to be worse than the box-type, and where the two approximate formulae for $\beta$ give widely differing results, are because the approximations have not here been corrected by Monte Carlos as in figure \ref{Neq1OnAxisBetafig}.
              }
         \label{SensyVsBoxSize}
   \end{figure}

The difference in size between the 1XMM and the optimized convolvers makes it more difficult to compare their efficiency. It seems a bit pointless to hobble the optimized convolver by restricting its size to the 1XMM $5\times5$ - after all, the whole aim of the exercise is to achieve the maximum practical improvement in sensitivity. It might be argued though that the discrepant sizes are unfair to the 1XMM `box' convolver - if bigger convolvers are sometimes better, might not much of the gain in going from a $5\times5$ box-type to a $9\times9$ optimized simply be due to the increase in size? It turns out however that box-type convolvers larger than $5\times5$ perform uniformly worse than the $5\times5$, as is shown in figure \ref{SensyVsBoxSize}; in fact, if we go in the other direction, to a $3\times3$ box, we get a slightly improved performance at all but the lowest background fluxes. Hence one can conclude that the comparison between convolvers of different size is probably being rather kind to the 1XMM algorithm than otherwise.

\begin{figure}
   \centering
      \resizebox{\hsize}{!}{\includegraphics[angle=-90]{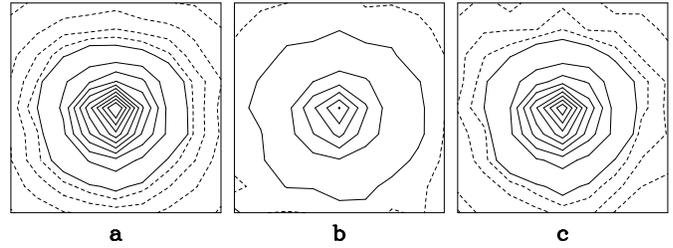}}
      \caption{Contour plots of the XMM-Newton PSF compared to two corresponding optimized convolution kernels. \textbf{a)} The on-axis PSF for the XMM-Newton EPIC PN camera at 1.25 keV, rebinned to 4 arcsec square pixels. \textbf{b)} Optimized convolution kernel for background $B=10^{-4}$ counts/pixel. \textbf{c)} Same, but optimized for background $B=10$ counts/pixel. All arrays were normalized before plotting. The main contour separation (solid contours) is 0.01; the lowest of these intervals has again been divided into five (dashed contours).
              }
         \label{Neq1OnAxisContourfig}
   \end{figure}

Some examples of optimized $9\times9$ kernels $w_{i,j}$ are compared to the PSF $S_{i,j}$ in figure \ref{Neq1OnAxisContourfig}. One would expect that $w \to 1$ as $B \to 0$ (all the counts are equally valuable) and $w \to S$ as $B \to \infty$ (Gaussian limit). This is consistent with the form of the high- and low-background kernels shown in the figure.

The quantity which should be compared is the counts amplitude $\beta_\mathrm{det}$ (as given for example in equation \ref{beta_def_ff}) at which the signal is just detectable - that is, at which the resulting null-hypothesis probability is just equal to some previously decided cutoff $P_\mathrm{det}$. Since the 1XMM sources were detected at a probability cutoff of $\exp(-8.0)$ (equivalent to about a 4.3-sigma detection), this is the value that was chosen for the present exercise.

\begin{figure}
   \centering
      \resizebox{\hsize}{!}{\includegraphics[angle=-90]{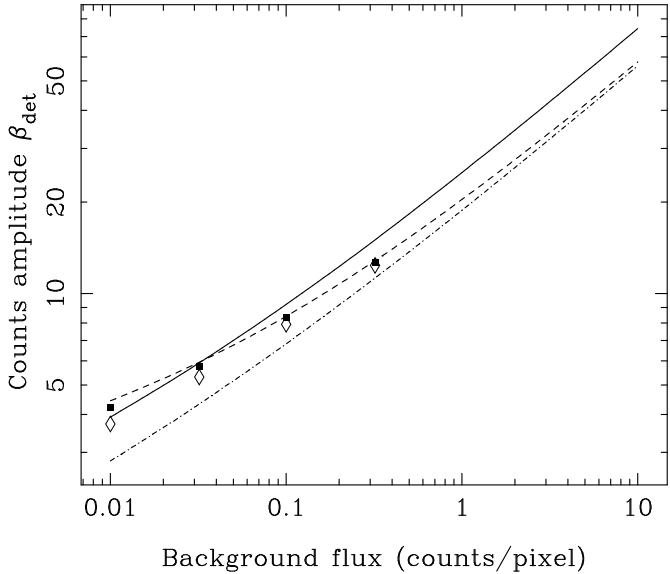}}
      \caption{The minimum detectable counts amplitude $\beta_\mathrm{det}$, plotted as a function of background flux $B$. The solid line represents the results obtained by convolving with a $5\times5$ unit array, as used in the creation of the 1XMM catalog. The remainder of the plot refers to a weighted-sum scheme in which the weights are optimized to the on-axis PSF of the XMM-Newton EPIC PN camera at 1.25 keV, binned up on a 9 by 9 grid of 4 arcsec square pixels. The dash-dot line shows the sensitivity predicted by the modified Fay and Feuer approximation formula ($\beta_\mathrm{det}$ given by equation \ref{beta_def_ff}). The dashed line gives the prediction of the $\chi^2$ formula ($\beta_\mathrm{det}$ given by equation \ref{beta_def_chi2}). The diamonds are samples of the `true' sensitivity of the matched filter method as derived from Monte Carlo experiments. The black squares, also Monte Carlo corrected, are the result of convolving with the PSF instead of the optimized convolver.
              }
         \label{Neq1OnAxisBetafig}
   \end{figure}

$\beta_\mathrm{det}$ as a function of background $B$ is plotted in figure \ref{Neq1OnAxisBetafig} for both the 1XMM and the matched-filter procedures. For the latter, since there is no exact formula for the null-probability distribution $P_\mathrm{null}$, the calculated sensitivity depends on the approximation used to represent $P_\mathrm{null}$. Shown on the figure are results (at finely-spaced values of background) of using respectively the Fay and Feuer (equation \ref{equ3}) and the $\chi^2$ (equation \ref{chi_for_many_W}) approximations, as well as four points (diamonds) where the true $P_\mathrm{null}$ has been estimated via Monte Carlos of $10^6$ iterations each.

As described in the introduction, several of the x-ray source detection procedures to be found in the literature include a step in which the raw images are convolved with the telescope PSF. The PSF is an optimum convolver in the Gaussian limit but may be expected to depart from the ideal at low values of background flux. To investigate this, the detection sensitivity obtained by use of the PSF as convolver was calculated at four levels of background. The resulting sensitivities, corrected via use of Monte Carlos to estimate $P_\mathrm{null}$, are plotted as black squares on fig \ref{Neq1OnAxisBetafig}.

Several conclusions can be drawn from figure \ref{Neq1OnAxisBetafig}. Firstly, for this form of signal, the Fay and Feuer approximation appears to be unserviceable for all but the highest values of background. The $\chi^2$ formula performs much better, accurately representing the true data down to about 0.2 counts/pixel background, below which it begins to diverge. Clearly though the matched-filter approach yields a better sensitivity than the $5\times5$ box convolver at all values of background, apparently asymptoting to about 25\% better at high background, the advantage decreasing to zero at low. The PSF appears to be a useful approximation to the optimum convolver for background levels greater than about 0.03 counts s$^{-1}$.

\begin{figure}
   \centering
      \resizebox{\hsize}{!}{\includegraphics[angle=-90]{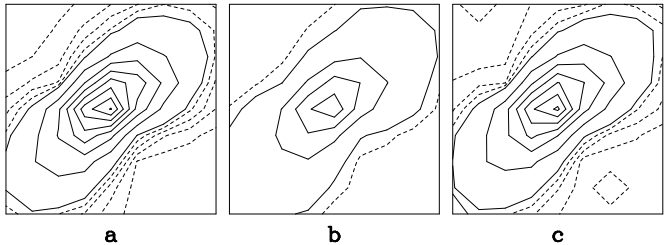}}
      \caption{Similar to figure \ref{Neq1OnAxisContourfig} except that the PSF at 850 arcsec from the edge of the optical axis and azimuth $=45 \degr$ was chosen.
              }
         \label{Neq1OffAxisContourfig}
   \end{figure}

\begin{figure}
   \centering
      \resizebox{\hsize}{!}{\includegraphics[angle=-90]{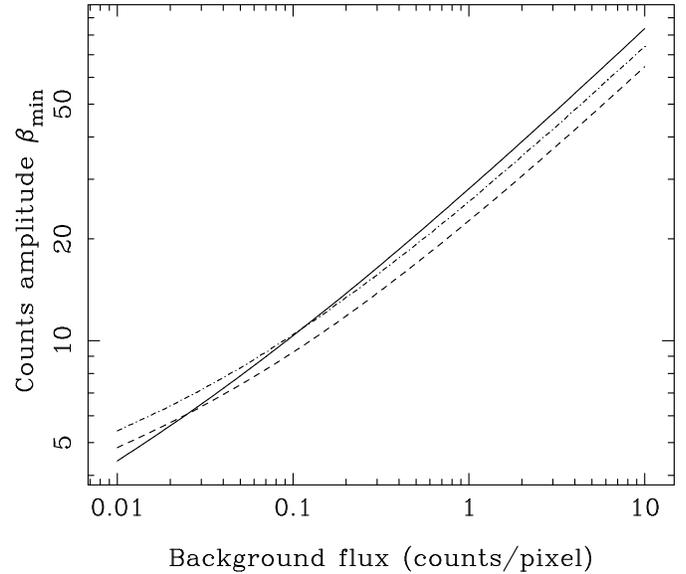}}
      \caption{Same as figure \ref{Neq1OnAxisBetafig} except that the PSF at 850 arcsec from the edge of the optical axis and azimuth $=45 \degr$ was chosen. The solid line again shows the results of the 1XMM method. Both the dashed and dot-dash lines give weighted-sum results, but the dashed line weights were optimized for the correct PSF for this field position whereas the weights for the dot-dash line were derived by optimizing for the on-axis (ie here incorrect) PSF.
              }
         \label{Neq1OffAxisBetafig}
   \end{figure}

The PSF of the XMM-Newton EPIC cameras becomes azimuthally distorted with distance from the centre of the field of view. It is therefore of interest to repeat the above exercise for an example of the off-axis PSF. In the present case, a PSF at 850 arcsec from the edge of the optical axis (94\% of the radius of the field-of-view of the EPIC cameras), at an azimuth of $45 \degr$, was arbitrarily chosen. All other variables were retained unchanged. No Monte Carlos were performed in this case, and only the results of the $\chi^2$ formula were used. The results can be seen in figures \ref{Neq1OffAxisContourfig} and \ref{Neq1OffAxisBetafig}.

Comparison between figures \ref{Neq1OnAxisBetafig} and \ref{Neq1OffAxisBetafig} shows that the degree of improvement to be gained through the use of matched filters is approximately constant across the field of view. Also, whatever the method used, the sensitivity decreases by about 15\% towards the edge of the field of view. This is because the less compact the PSF, the more difficult it is to sequester source from background counts - the effective background counts involved with the source are higher. For similar reasons, a decrease in sensitivity is seen in the detection of extended (ie non-pointlike) sources. Use of on-axis weights for the off-axis signal degrades the sensitivity by about 10\% over the whole range of background. For values of background lower than about 0.1 counts/pixel the sensitivity using this un-matched filter becomes nominally worse than that achievable via the box-convolver method, although no doubt some ground could be recovered by correction of the formula via Monte Carlos.

The variation in the shape of the PSF in XMM-Newton EPIC images puts some practical difficulties in the way of source detection via matched filters, because one has, in effect, to employ a different convolving kernel for each pixel of the image. This was the method adopted by Ueda et al (\cite{ueda}). An approximation to this which still allows one to use the useful mathematical properties of convolution is to divide the image into patches, convolve each patch separately, then add the results. Vikhlinin et al (\cite{vikhlinin}) used this technique. The XMM-Newton SAS (Gabriel et al \cite{carlos}) task \emph{asmooth} (Stewart \cite{asmooth}) can also perform such a piece-wise convolution.

%............................................................
\subsubsection{Matched filters for 5 energy bands}
\label{matched_at_5}

In this case it is no longer possible to invert the 1XMM and matched-filter methods in the same way, since the summed-likelihood approach used to find 1XMM sources in 5 bands (described in section \ref{1xmm_strategy}) cannot be expressed as a weighted sum of Poissonian integers. However, equations \ref{beta_def}, \ref{pnull_basic} and \ref{chi_null_prob} taken together amount to a relationship between the counts amplitude $\beta$ and the null-probability $P_\mathrm{null}$: hence one can numerically invert this relationship to obtain the sensitivity $\beta_\mathrm{det}$ which corresponds to the cutoff probability $P_\mathrm{det}$, which is left at $e^{-8}$ as before.

In order to calculate a matched filter for summing images in several energy bands, one must know the relative strength of the signal in each band, which amounts to knowing the source spectrum. Where sources have a variety of spectral shapes, as is the case for cosmic sources of x-rays, the matched-filter technique can only be optimized for a single class of sources at a time. The performance of the matched filter against an unmatched spectrum is examined toward the end of the present section.

For purposes of the present study, weights were optimized to detect x-ray sources with an absorbed power-law spectrum having a photon index 1.7 and a HI column density of $3.0 \times 10^{20}$ cm$^{-2}$. The relative count rates in each band were obtained, via the program \emph{xspec}, by folding this spectrum with the on-axis effective area function of the XMM-Newton PN camera. These weights are given in column 4 of table \ref{table:1}. The energy band definitions (columns 1 and 2) are those of 1XMM. The spectrum of the background (column 3), which is just as important for present purposes as that of the sources, was obtained from the 1XMM catalog in the following way: for each energy band, a 2-dimensional histogram was made of the background counts for each source versus the exposure time; the approximate minimum background count rate for the band was then estimated from this plot. To enable a check of the efficiency of the matched filter at detecting a source of spectrum far from nominal, weights derived from the EPIC PN count rates of 1XMM J175921.7-335322, one of the hardest sources in the 1XMM catalog, were also obtained: these are given in column 5 of the table.

   \begin{table}
   \caption{Source and background weights for the XMM-Newton PN camera.}             % title of Table
   \label{table:1}      % is used to refer this table in the text
   \centering                          % used for centering table
   \begin{tabular}{c c c c c}        % centered columns (5 columns)
   \hline\hline                 % inserts double horizontal lines
    \multicolumn{2}{c}{Band energies (keV)} & Background & Source & Hard\\ 
   Low & High & weights: & weights: & source:\\    % table heading 
   \hline                        % inserts single horizontal line
   0.2 &  0.5 & 0.0842 & 0.1208 & 0.0023 \\      % inserting body of the table
   0.5 &  2.0 & 0.1733 & 0.3763 & 0.0193 \\
   2.0 &  4.5 & 0.0941 & 0.0943 & 0.2993 \\
   4.5 &  7.5 & 0.0941 & 0.0350 & 0.6172 \\
   7.5 & 12.0 & 0.1485 & 0.0094 & 0.0619 \\ 
   \hline                                   %inserts single line
   \end{tabular}
   \end{table}

As noted in section \ref{matched_1_band}, a value of $N=4$ for the size of the optimized convolvers was chosen as giving the largest convolver which was still practical to compute. The optimization with 5 bands is observed to be a little slower than for the single-band case, understandable because there are are now $5 \times (2N+1)^2=405$ weights which must be optimized in parallel.

\begin{figure}
   \centering
      \resizebox{\hsize}{!}{\includegraphics[angle=-90]{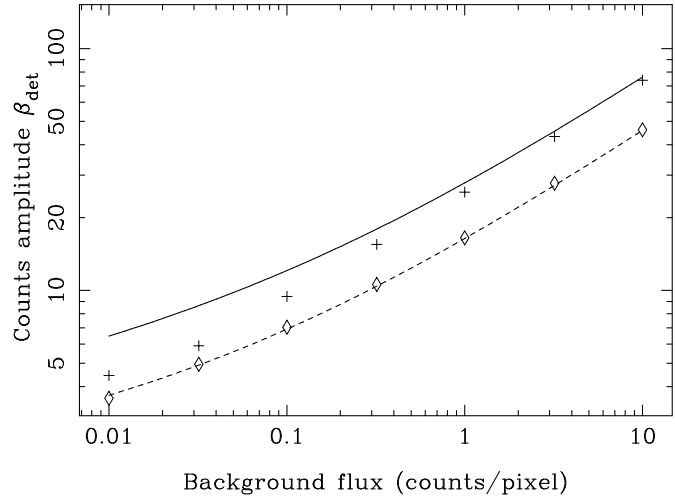}}
      \caption{This figure compares the theoretical sensitivity of two methods of detecting sources given several independent images of the same piece of sky. The methods compared here were designed to look for sources in XMM-Newton EPIC images made in 5 separate energy bands. The vertical scale shows the minimum detectable counts amplitude $\beta_\mathrm{det}$, whereas the horizontal scale gives the background counts per image pixel, summed over all energy bands. The solid line shows the nominal sensitivity of the source detection scheme used in the creation of the 1XMM catalog; the dashed line shows the nominal sensitivity of the matched-filter method ($\chi^2$ approximation). The crosses show values of 1XMM sensitivity corrected via a Monte Carlo and fitting approach. Diamonds show similarly corrected sensitivity values of the matched-filter method.
              }
         \label{Neq5OnAxisBetafig}
   \end{figure}

The two approaches are compared in figure \ref{Neq5OnAxisBetafig}. Before commenting on the difference in sensitivity between the two methods we ought to make sure we have an accurate measure of those sensitivities. As shown via a Monte Carlo experiment in section \ref{1xmm_strategy}, equation \ref{chi_null_prob} appears to be a poor approximation at a nett background level of 0.1 counts/pixel. The Monte Carlo exercise was repeated for six more background levels logarithmically spaced between 0.01 and 10 counts/pixel: that is, for each value of background, a Monte Carlo ensemble was generated, a cumulative distribution of the ensemble values $L$ was calculated, and finally a function $Q(\nu /2, L)$ was fitted to this distribution curve. The resulting values of $\nu$ are tabulated in table \ref{table:2}. The `canonical' value of $\nu$ in this case is 10 ($=2 \times 5$ bands).

   \begin{table}
   \caption{Values of $\nu$ obtained by fitting $Q(\nu /2,L)$ to Monte Carlo distributions of $L$ at different background levels.}             % title of Table
   \label{table:2}      % is used to refer this table in the text
   \centering                          % used for centering table
   \begin{tabular}{c c}        % centered columns (4 columns)
   \hline\hline                 % inserts double horizontal lines
   Nett background & Fitted\\ 
   (counts/pixel) & $\nu$ \\    % table heading 
   \hline                        % inserts single horizontal line
    0.01  & 4.31 \\      % inserting body of the table
    0.032 & 4.21 \\
    0.1   & 5.81 \\
    0.32  & 7.26 \\
    1.0   & 8.22 \\
    3.2   & 8.90 \\
   10.0   & 9.35 \\ 
   \hline                                   %inserts single line
   \end{tabular}
   \end{table}

Sensitivity values obtained by replacing the 5 in equation \ref{chi_null_prob} by the appropriate value of $\nu /2$ are shown by crosses in figure \ref{Neq5OnAxisBetafig}. Clearly, use of the unmodified formula exacts a significant sensitivity penalty (because of over-estimation of the rate of false positives) at values of background less than about 1.0 counts/pixel.

As was seen in section \ref{matched_1_band}, the $\chi^2$ formula used to approximate the null-probability distribution of weighted-Poisson-sum values also tends to diverge from the true distribution at low values of background. A similar set of Monte Carlo corrections was therefore performed for the weighted-sum data. Fortuitously, for the 5-band signal presently chosen, the corrections appear to be insignificant. The reader is cautioned not to expect this to be the case for all signal shapes.

\begin{figure}
   \centering
      \resizebox{\hsize}{!}{\includegraphics[angle=-90]{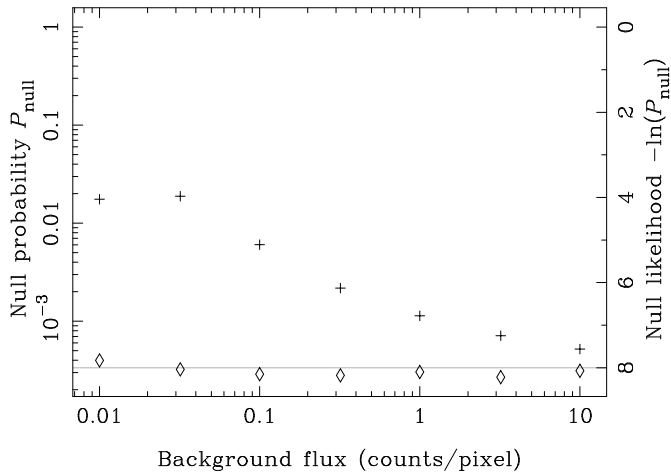}}
      \caption{Nominal values of detection likelihood which should be chosen in order to obtain a true detection likelihood of 8. The crosses show the results relevant to the 1XMM procedure; diamonds refer to the matched-filter method ($\chi^2$ formula).
              }
         \label{likeMinDeviation}
   \end{figure}

As mentioned in section \ref{many_n_theory}, one may compensate for the distortion in the true null likelihood (thus in the rate of false positives due to background fluctuations) by changing the value of nominal null likelihood used to sort `sources' from `non-sources'. Such compensating values of null likelihood for seven values of background, evenly-spaced on a logarithmic scale, are plotted in figure \ref{likeMinDeviation}.

Taking the corrections into account, at high background values the matched-filter method appears to offer about 1.6 times the sensitivity of the 1XMM procedure, the advantage decreasing gradually to about 1.2 at the lowest background value plotted.

A hasty comparison of figures \ref{Neq1OnAxisBetafig} and \ref{Neq5OnAxisBetafig} may lead one to conclude that there is, paradoxically, not much advantage to be gained by detecting sources over 5 bands rather than 1. However, the vertical scales of the two graphs are not equivalent, because they refer to different signal shapes. To make a proper comparison one would need to multiply all the background values of figure \ref{Neq5OnAxisBetafig} by 0.1733 (the proportion of the total background found in energy band 2) and, for like reasons, all the amplitude values by 0.3763. Comparison with figure \ref{Neq1OnAxisBetafig} then reveals an improvement in sensitivity for 5- versus 1-band detection by about a factor of 2 for the matched-filter method and 1.5 for the 1XMM method.

\begin{figure}
   \centering
      \resizebox{\hsize}{!}{\includegraphics[angle=-90]{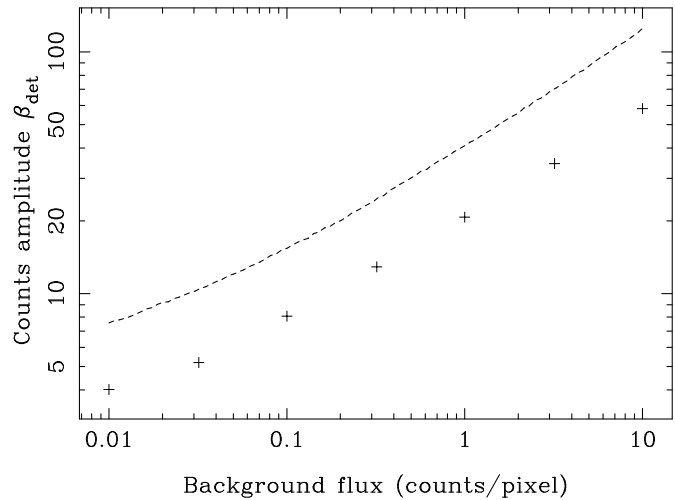}}
      \caption{Same as figure \ref{Neq5OnAxisBetafig}, except that the sensitivity values were calculated using the hard-source weights of column 5 of table \ref{table:1}. Only the Monte-Carlo-corrected points for the 1XMM procedure (crosses) and the $\chi^2$-theory curve for the matched-filter procedure (dashed line) are shown.
              }
         \label{Neq5OnAxisBetaHardfig}
   \end{figure}

As mentioned earlier, a filter which has been optimized to detect sources of a particular spectrum may not perform well in the detection of sources with very different spectra. To investigate this, the exercise of figure \ref{Neq5OnAxisBetafig} was repeated. The filter was optimized as before for a source having weights listed in column 4 of table \ref{table:1}, but the sensitivity values were calculated under the assumption that the source weights came from column 5. The results are plotted in figure \ref{Neq5OnAxisBetaHardfig}.

It is apparent that the sensitivity of the matched filter to the hard-spectrum source is significantly degraded - indeed by a factor of 2 - whereas the detection efficiency of the 1XMM procedure is if anything slightly improved. Although this hard spectrum is the worst case likely to be encountered in practice, the results suggest the desirability of using a non-matched procedure in parallel. It is however also worth noting in passing that, regardless of whether the detection technique is matched to a particular source spectrum or not, some spectrum must be assumed in order to calculate any kind of multi-band sensitivity value.

%............................................................
\subsubsection{A check on the sensitivity results}

\begin{figure}
   \centering
      \resizebox{\hsize}{!}{\includegraphics[angle=-90]{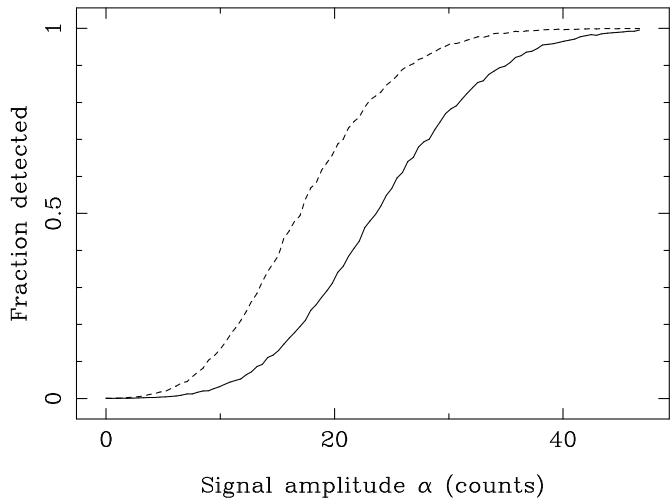}}
      \caption{Detection sensitivities of the 1XMM (solid line) and matched-filter (dashed line) methods are compared via a Monte Carlo experiment. The vertical scale gives the fraction of `sources' at each value of $\alpha$ which were `detected'.
              }
         \label{forwardMonte}
   \end{figure}

So far we have been `working backwards', inverting expressions to obtain estimates of the minimum signal detectable under a variety of conditions. As a check on this procedure, a `forwards' Monte Carlo experiment was performed as follows. Random data were generated from parent distributions of the form of equation \ref{equ0} for a sequence of values of the amplitude $\alpha$. The data were generated in 5 bands, using the PN background ratios tabulated in table \ref{table:1}, with a net background flux of 1.0 counts/pixel. The usual PSFs provided the signal appropriate to each band. An ensemble of $10^4$ mini images was accumulated at each of 200 equally-spaced values of alpha. Each 5-band image stack was submitted to both the 1XMM and the matched-filter source detection procedures. The detection frequencies as functions of $\alpha$ are compared in figure \ref{forwardMonte}. From figure \ref{Neq5OnAxisBetafig}, one would expect that signals of $\alpha >$ about 16 would be detected by the matched-filter approach, as opposed to a cutoff of about 25 for the (uncorrected) 1XMM approach. Although statistical fluctuations mean that signals with $\alpha < \beta_\mathrm{det}$ for that background flux are occasionally detected, and signals with $\alpha > \beta_\mathrm{det}$ are occasionally not, the results of this Monte Carlo are consistent with the earlier analysis.

%++++++++++++++++++++++++++++++++++++++++++++++++++++++++++++
\section{Conclusions}

It is not possible to detect with certainty a signal superimposed on a noisy background where there is some non-zero probability that a combination of random background values can mimic the signal: the best that can be done is to calculate a detection probability - or its complement, the null or non-detection probability. The aim of any method to enhance signal detectability is therefore to decrease the null probability of a signal of any given amplitude, or, equivalently, to decrease the amplitude at which a signal generates a given value of null probability. This must be achieved by performing some transform upon the set of measured values of signal plus background plus noise.

The main thrust of the present paper has been to examine that subset of transforms which may be expressed as discrete convolutions of the input data. There appears to be nothing new to say about the case in which the probability distribution of the noise value in any given data channel is Gaussian; because of this, attention has been further restricted to the Poissonian regime. Two independent approximations to the null-probability distribution for a convolution of Poissonian data have been described and compared; such approximate formulas allow one to optimize the convolution for detection of signals of any specified shape without performing a Monte Carlo on each occasion.

Although the method described here yields optimum signal detection via convolution, the theory says nothing about the possibilities or otherwise of obtaining better detection sensitivity via other, non-linear transforms. Despite this, the optimized-convolution method has been shown to perform substantially better over a wide range of background levels than the more complicated technique which was employed to detect, in XMM-Newton x-ray images, the sources which comprise the 1XMM catalog. However this is perhaps not surprising in view of the fact that the convolution method makes use of more information about the signal (in the 1XMM case `information' means the average spectrum and point-spread function (PSF) of the x-ray sources) than the 1XMM method.

The greater use of information about the signal exposes one of the drawbacks of the optimized-convolver method: namely, that if the shape of the signal is not known, or can vary, then assumptions must be made about it. Any given convolver is optimized for only one shape of signal (and for one particular level of background). To detect with optimal sensitivity signals having a variety of shapes, under a variety of background conditions, one would need to repeat the convolution with many different convolvers, each tailored to a different signal/background combination. In practice this may not be worth the extra effort, since in many cases it may happen that acceptable results can be obtained by using a relatively small set of convolvers. Taking the XMM-Newton x-ray images as an example, it was shown in section \ref{matched_1_band} that use of the on-axis PSF across the whole field results in only a few percent loss of sensitivity even towards the edges of the field of view where the PSF becomes significantly elongated in the azimuthal direction. In addition, the form of the optimal convolver appears to be relatively insensitive to background level. As regards the x-ray spectrum, it should be remembered that the weights tabulated in table \ref{table:1} represent x-ray spectra after convolution by the strongly peaked response function of the instrument, and their trend is therefore dominated by that response function. Sources with quite different spectra may thus be expected to yield similar sets of spectral weights; for example, the bulk of x-ray sources in 1XMM exhibit their highest and lowest fluxes in bands 2 and 5 respectively. However it is probably desirable in practice to supplement the full matched-filter procedure with a non-spectrum-specific detection algorithm. The 1XMM sliding-box method, perhaps with the `boxes' replaced by PSF-matched convolvers for increased sensitivity, is a possible choice for the latter.

Finally, some discrimination between signals may be desirable, since not all signal shapes represent sources we would wish to detect. In the x-ray case, whereas the 1XMM detection method cannot discriminate between an x-ray source and a bright pixel on the CCD\footnote{a further \emph{characterisation} step in the 1XMM chain, employing the SAS task \emph{emldetect}, does however (in principle) allow the two to be discriminated.}, the matched-filter method does offer some degree of selection against bright pixels and other artifacts.

Another caveat to be mentioned is the fact that both the formulas presented for null probability of a weighted sum of Poissonian data are only approximate; and what is perhaps worse, no analysis has yet been presented which would allow one to estimate the goodness of the approximations. In this case one can only fall back on probability curves derived from Monte Carlo data with which to calibrate the approximation formulae. A good example of the desirability of checks of this kind is the large gap demonstrated between the null-probability approximation used for 1XMM source detection and the results of Monte Carlos at a variety of levels of background. Although the approximations for the matched-filter method do not appear to be nearly so unsatisfactory, they are not immune from difficulties of this sort and should be subject to similar checks in practice.

A natural extension of the convolver method as applied to x-ray source detection is to allow one to correctly add together overlapping images. One's first impulse in this situation is simply to add the images without weighting, but (certainly in the XMM-Newton case) because the background rate can vary with time, separate images may have different source/background ratios and should therefore be weighted accordingly. The theory described in this paper allows one to estimate the optimum weights for such combinations.

The author hopes to make the matched-filter method available in the \emph{edetect} package of the v-6.5 release of the XMM SAS.

\appendix
\section{The x-ray background} \label{confusion}

As mentioned in the introduction, there is good evidence that the bulk of the cosmological x-ray background consists of numerous faint discrete sources. The present paper assumes however that the x-ray sky may be modelled by sparse, relatively bright sources (though not all necessarily detectable at any given exposure duration) upon a relatively smooth background. It is therefore of interest to see how consistent this model is with our current understanding of the real sky.

The fundamental quantity to deal with is the probability distribution of detected flux, where the ensemble is taken to consist of measurements at random directions in the sky; non-source background is neglected and it is assumed that each measurement is free from other sorts of variation, eg the Poissonian detection noise we have been discussing so far. The connection between the distribution of source flux and the distribution of detected source counts depends in a mathematically complicated way on the `beam shape', or PSF in our case. Useful treatments of the topic can be found in Condon (\cite{condon}) and Scheuer (\cite{scheuer}).

For present purposes it is enough to consider the standard deviation of such a probability distribution. Let us consider the case in which the distribution of source flux $n(S)$ obeys a power law, viz:

\begin{displaymath}
  n(S) = kS^{-\gamma}.
\end{displaymath}
Condon derives the following formula for standard deviation $\sigma$ in this case:\footnote{Actually Condon's equation (13) appears to be incorrect: according to my working, the $D^{3-\gamma}_\mathrm{c}$ should be inside the square root. In addition, the condition $\gamma>2$, although necessary for others of his results, is not required here.}

\begin{equation} \label{conf_sigma}
  \sigma = \left( \frac{k\Omega_\mathrm{e}C_\mathrm{max}^{3-\gamma}}{3-\gamma} \right)^{1/2},
\end{equation}
where $C_\mathrm{max}$ is some upper limit on the detected source counts and $\Omega_\mathrm{e}$, the equivalent solid angle of the PSF, is defined as

\begin{equation} \label{equiv_omega}
  \Omega_\mathrm{e} = \int \left[ psf(\theta,\phi) \right]^{\gamma-1} \, d\Omega.
\end{equation}
Note that $\sigma$ is unbounded for the entire probability distribution (ie, as $C_\mathrm{max} \to \infty$).

As for $n(S)$, the measurements described in Mushotzky et al (\cite{mushotzky}) are consistent with a `two-slope' model in which, for energies from 0.5 to 2 keV (corresponding to energy band 2 of 1XMM), $\gamma$ has the value 1.7 below $S=7 \times 10^{-15}$ erg cm$^{-2}$ s$^{-1}$ and 2.5 above it. The $k$ values can be evaluated from

\begin{displaymath}
  \frac{k_\mathrm{faint}}{1-\gamma_\mathrm{faint}}S^{1-\gamma_\mathrm{faint}} = \frac{k_\mathrm{bright}}{1-\gamma_\mathrm{bright}}S^{1-\gamma_\mathrm{bright}} \sim 200 \, \mathrm{deg}^{-2}
\end{displaymath}
where $S$ is the flux at the `knee'.

Which $\gamma$ to choose? 200 sources per square degree, the integrated number density at the knee, yields about 30 sources per EPIC PN field. This is quite a typical number of serendipitous sources to find in such fields: hence one can say that the PN camera will, in a typical exposure, detect sources fainter than the flux at the knee. I therefore choose a value of 1.7 for $\gamma$. Substitution of this value into equation \ref{equiv_omega} yields $2.34 \times 10^{-5}$ deg$^2$ for the equivalent area of the on-axis, band 2 EPIC PN PSF.

At this point it is convenient to convert fluxes $S$ to counts $C$. The highest value of background considered in the present paper is 10 counts pixel$^{-1}$. By use of the same histogram technique as described in section \ref{matched_at_5}, one may deduce that the typical background count rate for EPIC PN in 1XMM band 2 is about $2 \times 10^{-5}$ counts pixel$^{-1}$ s$^{-1}$. A background value of 10 counts pixel$^{-1}$ thus corresponds to an exposure time of $5 \times 10^5$ s, which is 5 times longer than the longest (PN) exposure time in the 1XMM catalog. Some multi-epoch observations made with XMM-Newton may approach this duration however; hence I take it as a reasonable upper limit to practical XMM-Newton observations. The flux to count-rate conversion factor for 1XMM band 2 (calculated in the same exercise as the source weights tabulated in table \ref{table:1}) is $7.50 \times 10^{11}$ counts erg$^{-1}$ cm$^2$; the `knee' in the Mushotzky soft-band log$N$-log$S$ diagram thus falls, for a $5 \times 10^5$ s exposure time, at $\sim$2600 counts in 1XMM band 2 and (making use of the source weights in table \ref{table:1}) 7000 counts in the total band. Comparison of these numbers with figure \ref{Neq5OnAxisBetafig} shows that even the 1XMM algorithm could detect sources 2 orders of magnitude fainter than the `knee' at this exposure duration; clearly the estimate in the previous paragraph that XMM-Newton is capable of seeing far past the `knee' in the log$N$-log$S$ curve is correct.

One finds that $k_\mathrm{faint}$ in these units works out to be $6.88 \times 10^4$. We are now in a position to use equation \ref{conf_sigma} to calculate the `standard deviation' $\sigma$ of the noise in the ensemble of measurements. Equation \ref{conf_sigma} evaluates to

\begin{displaymath}
  \sigma = 1.11 \times C_\mathrm{max}^{0.65}.
\end{displaymath}
It only remains to select a value for $C_\mathrm{max}$. Suppose we choose $C_\mathrm{max}=47$ counts, which from figure \ref{Neq5OnAxisBetafig} is the detection sensitivity obtainable (under the assumption that the background is flat) at this exposure length using the matched-filter algorithm described in the present paper. $\sigma$ evaluates to $\sim$13 counts, which is several times larger than the standard deviation $\sqrt{10}$ of the Poisson noise. However, the log$N$-log$S$ model indicates that the total source density at this counts value is 6640 deg$^{-2}$, which is still only 0.16 sources per PSF equivalent area $\Omega_\mathrm{e}$. The counts value at which one expects 1 source in total per $\Omega_\mathrm{e}$ is 3.3; using this for $C_\mathrm{max}$ yields 2.4 counts for $\sigma$ instead, just lower than the Poisson noise. Thus we may conclude that, for XMM-Newton exposures of total duration up to $5 \times 10^5$ s in length, the x-ray sky may still be modelled to acceptable accuracy by a flat background with superposed sources. It is clear though that the next generation of x-ray telescopes may not have life so easy.

Finally, a word about background estimation. So far in the present paper it has been assumed that the background contribution is known. Background can be difficult to estimate, though - to calculate the background one must first excise or avoid the sources, but it is difficult to find sources without first having a knowledge of the background. However, one can conceive of an iterative process in which the detectable sources are gradually detected and excised in parallel with improvements in the background estimate. In the present case that would still leave a population of sources which are too faint to be detected but which are brighter than the confusion level calculated in the preceding paragraph. The sum of these faint sources will bias the background estimate upward. The average counts deg$^{-2}$ contributed by these sources is

\begin{displaymath}
  \phi = \int_{C_\mathrm{conf}}^{C_\mathrm{det}} dC \ C \ n(C).
\end{displaymath}
For a single-power-law $n=kC^{-\gamma}$ this gives
\begin{displaymath}
  \phi = \frac{k}{2-\gamma} \left( C_\mathrm{det}^{2-\gamma} - C_\mathrm{conf}^{2-\gamma} \right).
\end{displaymath}
If we use $6.88 \times 10^4$ for $k$, 1.7 for $\gamma$, 47 for $C_\mathrm{det}$ and 3.3 for $C_\mathrm{conf}$, $\phi$ evaluates to $4 \times 10^5$ counts deg$^{-2}$, or 0.5 counts pixel$^{-1}$ for 4 arcsec square pixels. A similar calculation can of course be performed for any other exposure duration.

\begin{acknowledgements}
I would like to thank Silvia Mateos for calculating the source weights in table \ref{table:1}.
\end{acknowledgements}

\end{document}